\documentclass[conference,a4paper]{IEEEtran}

\usepackage[utf8]{inputenc} 
\usepackage[T1]{fontenc}
\usepackage{url}
\usepackage{ifthen}
\usepackage{cite}
\usepackage[cmex10]{amsmath} %

\usepackage{amssymb}
\usepackage{aligned-overset}

\usepackage{mleftright}       %
\mleftright                   %

\usepackage{graphicx}         %

\usepackage[caption=false,font=footnotesize]{subfig}

\usepackage{tikz}
\usetikzlibrary{arrows.meta}
\usetikzlibrary{decorations.pathreplacing,calligraphy}

\usepackage{pgfplots}
\pgfplotsset{compat=1.18}

\usepackage{nicematrix}

\usepackage[hidelinks]{hyperref}
\usepackage[capitalise]{cleveref}
\crefname{construction}{Construction}{Constructions}
\crefname{prop}{Proposition}{Propositions}

\newtheorem{theorem}{Theorem}%
\newtheorem{lemma}{Lemma}%

\newtheorem{definition}{Definition}%
\newtheorem{construction}{Construction}%
\newtheorem{example}{Example}%

\definecolor{violettblau}{cmyk}{0.9, 0.6, 0, 0}
\definecolor{rot}{RGB}{238, 28 35}

\newcommand{\RvE}{\mathsf{E}}

\newcommand{\RvK}{\mathsf{K}}

\newcommand{\RvR}{\mathsf{R}}

\newcommand{\RvU}{\mathsf{U}}

\newcommand{\RvX}{\mathsf{X}}

\newcommand{\vA}{\mathbf{A}}

\newcommand{\vI}{\mathbf{I}}

\newcommand{\vL}{\mathbf{L}}
\newcommand{\vM}{\mathbf{M}}

\newcommand{\vT}{\mathbf{T}}

\newcommand{\vV}{\mathbf{V}}

\newcommand{\va}{\mathbf{a}}
\newcommand{\vb}{\mathbf{b}}
\newcommand{\vc}{\mathbf{c}}

\newcommand{\ve}{\mathbf{e}}
\newcommand{\vf}{\mathbf{f}}

\newcommand{\vm}{\mathbf{m}}

\newcommand{\vp}{\mathbf{p}}

\newcommand{\vr}{\mathbf{r}}

\newcommand{\vu}{\mathbf{u}}

\newcommand{\vell}{\boldsymbol{\ell}}
\newcommand{\vbeta}{\boldsymbol{\beta}}

\newcommand{\trans}{^{\mathsf{T}}}

\newcommand{\fiel}{\mathbb{F}}
\newcommand{\fielz}{\mathbb{F}^*}
\newcommand{\Fq}{\mathbb{F}_q}
\newcommand{\Fqm}{\mathbb{F}_{q^m}}
\newcommand{\Fqmx}{\mathbb{F}_{q^m}[x;\sigma]}

\newcommand{\natz}{\mathbb{N}_0}
\newcommand{\nat}{\mathbb{N}}

\newcommand{\Vand}{\vV^{\sigma}}

\newcommand{\Ero}{\mathcal{E}_1}
\newcommand{\Ert}{\mathcal{E}_2}

\newcommand{\st}{\mathrm{st}}
\newcommand{\dl}{\mathrm{dl}}

\newcommand{\Cod}{\mathcal{C}}

\newcommand{\Clrs}{\Cod_{\mathrm{LRS}}^{\sigma, k}}
\newcommand{\Cout}{\Cod_{\mathrm{out}}}

\newcommand{\Cloci}{\Cod_{\mathrm{loc},i}}
\newcommand{\Cglob}{\Cod_{\mathrm{glob}}}

\newcommand{\out}{{}_{\mathrm{out}}}
\newcommand{\loc}{{}_{\mathrm{loc}}}
\newcommand{\glob}{{}_{\mathrm{glob}}}

\newcommand{\Delglob}{\Delta_{\mathrm{gl}}}
\newcommand{\Delglobone}{\Delta_{\mathrm{gl},1}}
\newcommand{\dir}{{}_{,\mathrm{dir}}}
\newcommand{\for}{{}_{,\mathrm{fw}}}

\newcommand{\ks}{k_{\mathrm{s}}}

\newcommand{\hs}{{}_{\mathrm{s}}}
\newcommand{\he}{{}_{\mathrm{e}}}
\newcommand{\ke}{k_{\mathrm{e}}}

\newcommand{\FL}{\mathcal{F}}
\newcommand{\Gtwo}{\mathcal{G}_{l_2}}

\newcommand{\FLbar}{\mathcal{F}^\prime}

\newcommand{\upn}[1]{{}_{\mathrm{up},{#1}}}
\newcommand{\upi}{{}_{\mathrm{up},i}}

\newcommand{\phiinv}{\varphi^{-1}}

\newcommand{\bell}{\boldsymbol{\ell}}

\DeclareMathOperator{\MI}{I}
\DeclareMathOperator{\Ent}{H}
\DeclareMathOperator{\diag}{diag}
\DeclareMathOperator{\rk}{rk}
\DeclareMathOperator{\Px}{P}

\begin{document}
\title{Secure Storage using Maximally Recoverable Locally Repairable Codes}

\author{
  \IEEEauthorblockN{Tim Janz}
  \IEEEauthorblockA{Institute of Telecommunications\\
                    University of Stuttgart, Germany\\
                    tim.janz@inue.uni-stuttgart.de}
  \and
  \IEEEauthorblockN{Hedongliang Liu, Rawad Bitar}
  \IEEEauthorblockA{Institute of Communications Engineering\\ 
                    Technical University of Munich, Germany\\
                    \{lia.liu, rawad.bitar\}@tum.de}
  \and
  \IEEEauthorblockN{Frank R.~Kschischang}
  \IEEEauthorblockA{Edward S. Rogers Sr. Department of\\ Electrical \& Computer Engineering\\ 
                    University of Toronto, Canada\\
                    frank@ece.utoronto.ca}
}

\maketitle

\begin{abstract}
  This paper considers data secrecy in distributed storage systems (DSSs) using maximally recoverable locally repairable codes (MR-LRCs).
  Conventional MR-LRCs are in general not secure against eavesdroppers who can observe the transmitted data during a global repair operation. This work enables nonzero secrecy dimension of DSSs encoded by MR-LRCs through a new repair framework. 
  The key idea is to associate each local group with a central processing unit (CPU), which aggregates and transmits the contribution from the intact nodes of their group to the CPU of a group needing a global repair. 
  The aggregation is enabled by so-called local polynomials that can be generated independently in each group.
  Two different schemes -- direct repair and forwarded repair -- are considered, and their secrecy dimension using MR-LRCs is derived. Positive secrecy dimension is enabled for several parameter regimes.
  \end{abstract}

\section{Introduction}\label{sec:intro}%
Locally repairable codes (LRCs) are used in distributed storage systems (DSSs) to protect data against loss due to node failures \cite{gopalan2012on,huang2007pyramid,papailiopoulos2014locally,kamath2014codes}. A class of LRCs introduced in \cite{gopalan2014explicit}, called \emph{maximally recoverable locally repairable} codes (MR-LRCs~\cite{chen2007maximally,martinez-penas2019universal,gopi2020maximally,guruswami2020constructions,cai2022construction,gopi2022improved}, also known as \emph{partial maximum-distance separable} (PMDS) codes \cite{blaum2013partial,calis2016general,gabrys2018constructions,neri2020random,bogart2021constructing}), can correct any erasure pattern that is information-theoretically correctable for a specified level of redundancy. 
In a DSS encoded by an MR-LRC,
nodes are partitioned into local groups. Each group can repair a certain number of node failures locally,
while additional failures can be also repaired by a so-called \emph{global repair}.
In this work, we consider DSSs
encoded by MR-LRCs of block length $N$ with $g$ local groups, locality $r$, local distance $\delta$ and $h=N-gr$ global parities.   This means that each group can repair up to $\delta-1$ node failures by contacting $r$ intact nodes within the same group.  The $h$ global parities permit repair of up to $h$ additional failed nodes at arbitrary locations.
  
A desirable feature of DSSs is \emph{secrecy}, i.e., the stored information should be kept secret even if some nodes are monitored by an eavesdropper.
An LRC construction that maintains secrecy in the presence of a so-called \emph{$(l_1,l_2)$-eavesdropper} was proposed in \cite{rawat2014optimal}.
We assume a similar, though slightly altered, eavesdropper model where the eavesdropper has read-access to any $l_1$ nodes and, in addition, can
observe all data traffic to---and the stored data within all nodes of---any $l_2$ local groups.
In a DSS storing $k$ independent symbols and having locality $r$, were an eavesdropper to observe any $k$ independent symbols, it would be able to reconstruct the whole DSS, resulting in no secrecy.
Thus, to construct a DSS with a positive secrecy dimension, we assume throughout this paper that $l_1+l_2r<k$.
 
\begin{figure} [t]
      \centering   
      \begin{tikzpicture}[
      square/.style = {draw, rectangle, minimum size=\m, outer sep=0, inner sep=0, font=\small},
                              ]
      \def\m{20pt}
      \def\off{25pt}
      \def\circoff{6pt}
      \def\trioff{7pt}
      \def\tri{6pt}
      \def\circsize{2pt}
      \def\w{5}
      \def\h{3}
      \def\loc{3}

        \foreach \x in {1,...,\w}
          \foreach \y in {1,...,\h}
             {  
             \pgfmathtruncatemacro{\xin}{\x-1}
                \ifnum\x>\loc
                    \ifnum\y<\h
                        \node [square, fill=gray!30]  (\x_\y) at (\x*\m,-\y*\off) {$x^{(\xin)}_{\y} $};
                    \else
                        \node [square, fill=gray!30]  (\x_\y) at (\x*\m,-\y*\off) {$ x^{(\xin)}_{\y} $};
                    \fi
                 \else
                    \ifnum\y>2
                        \node [square, fill=gray!70]  (\x_\y) at (\x*\m,-\y*\off) {$ x^{(\xin)}_{\y} $};
                    \else
                        \node [square, fill=white]  (\x_\y) at (\x*\m,-\y*\off) {$ x^{(\xin)}_{\y} $};
                    \fi
                \fi
             }

      \node [square, fill=gray!30]  (b) at (4*\m,-1*\off) {$\blacklozenge$};
      \node [square, fill=gray!30]  (c) at (5*\m,-1*\off) {$\blacklozenge$};
  
      \node [square, fill=white]  (d) at (2*\m,-2*\off) {$\blacklozenge$};
      \node [square, fill=white]  (e) at (3*\m,-2*\off) {$\blacklozenge$};
  
      \node [square, fill=white]  (f) at (1*\m,-3*\off) {$\blacklozenge$};
      \node [square, fill=gray!30]  (g) at (5*\m,-3*\off) {$\blacklozenge$};

      \node [square, fill=white]  (l) at (3*\m,-1*\off) {$\bigstar$};
      \node [square, fill=white]  (l) at (1*\m,-1*\off) {$\bigstar$};

      \filldraw[violettblau] (1*\m-\circoff,-2*\off+\circoff) circle (\circsize);
      \filldraw[rot] (1*\m-\trioff-\tri,-1*\off+\trioff) -- (1*\m-\trioff,-1*\off+\trioff) -- (1*\m-\trioff-\tri/2,-1*\off+\trioff+\tri) --(1*\m-\trioff-\tri,-1*\off+\trioff);
        
      \end{tikzpicture}
      \caption{Illustration of a DSS with $N = 15$ nodes and storing $k=7$ independent symbols. 
      The DSS is encoded by an MR-LRC with $g=3$ groups, locality $r=3$, local distance $\delta=3$ (parities in light gray) and $h=2$ global parities (in dark gray). 
      The failed nodes marked with diamonds can be repaired locally while the failed nodes marked with stars need data from other groups to be repaired.
      The DSS is in presence of a  $(1,1)$-eavesdropper who can read the data stored on one node (marked by a blue circle) and the downloaded and stored data of any node in the top group (marked by a red triangle).
    } 
      \label{fig:MR-LRC_example}
  \end{figure}
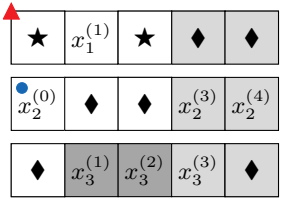

In this work, we study the secrecy of DSSs encoded with MR-LRCs.
 \cref{fig:MR-LRC_example} illustrates a DSS that is encoded with an MR-LRC in the presence of an $(l_1,l_2)$-eavesdropper. 
 We define the \emph{secrecy dimension} as the minimum number of independent information symbols about which
  an $(l_1,l_2)$-eavesdropper cannot gain any information.
  The secrecy dimension of an LRC-equipped DSS has been characterized in \cite{rawat2014optimal,agarwal2016security}. 
With the ability to globally repair failed nodes, DSSs encoded by MR-LRCs gain higher reliability than DSSs with LRCs. 
However, in the presence of an $(l_1,l_2)$-eavesdropper with $l_2 > 0$,
the secrecy of the DSS comes under threat, possibly leading to zero
secrecy dimension, since the global repair process enables the eavesdropper to obtain information from other groups.

The main contributions of this work are two new global repair schemes for DSSs encoded with MR-LRCs and the characterization of the secrecy dimensions achieved by these schemes.
The main idea enabling a nonzero secrecy dimension is the introduction of a central processing unit (CPU) for each local group that serves as the interface 
for all communication with other local groups during a global repair.
  When a global repair is required, the CPU in each group having intact nodes uses so-called \emph{local polynomials} to generate a symbol that is the contribution of that group to the global repair.
  The respective CPU then sends the group's contribution to the other CPUs according to the particular information-sharing scheme.
The derived secrecy dimensions show that the DSSs, encoded with MR-LRCs and equipped with a global or forwarding global repair scheme, can achieve positive secrecy dimensions in presence of an $(l_1,l_2)$-eavesdropper.

  \section{Preliminaries}
  \label{sec:prelim}
    For integers $a$, $b$ with $a\leq b$, let $[a,b]$ denote the set $\{a,a+1,\ldots,b-1,b\}$ and for $b \geq 1$, let $[b]$ denote the set $[1,b]$.
  The set of positive integers is denoted as $\nat$ and $\natz:=\nat\cup \{0\}$. Let $\Fqm$ denote a finite extension field of degree $m$ with base field $\Fq$, where $q$ is a prime power. When the size is not relevant, we simply write $\fiel$ and we let $\fielz= \fiel\backslash \{0\}$.  The rank of a matrix $\mathbf{M}$ with entries from $\fiel$ is denoted as $\rk(\mathbf{M})$.
  The Hadamard (element-wise) product of two vectors $\mathbf{u}, \mathbf{v} \in \fiel^n$ is denoted as $\mathbf{u} \odot \mathbf{v}$.
  The entropy of a discrete random variable $\RvX\in \mathcal{X}$ is defined as 
  $\Ent(\RvX)=-\sum_{a\in\mathcal{X}:\Px_{\RvX}(a)>0} \Px_{\RvX}(a)\log_{|\mathcal{X}|}\Px_{\RvX}(a)$. 
  Throughout this paper, two different node indexings are used. The first is a DSS motivated indexing with group index $i\in[g]$ and node index $j\in [0,r+\delta-2]$, e.g., in \cref{fig:MR-LRC_example}, the second node in the third group is denoted by $x_3^{(1)}$. 
  The second indexing is only using one index $\mu\in[0,N-1]$ and there is a bijective mapping $\varphi:\natz\times \nat \rightarrow \natz$ with $(i,j)\mapsto \mu = j+(r+\delta -2)(i-1)$. 
  The inverse mapping $\phiinv$ is written as $\phiinv(\mu)=(i(\mu),j(\mu))$, where $i(\mu)= \lceil \frac{\mu}{r+\delta -2}\rceil$ and $j(\mu)=\mu \mod (r+\delta -2)$.

  \subsection{Linearized Reed--Solomon Codes}
  \label{subsec:skew-codes}

  Let $\Fqm[x;\sigma]$ denote the ring of skew polynomials with automorphism $\sigma: \Fqm \rightarrow \Fqm$ such that $\sigma(a)=a^{q}$ for $a\in \Fqm$.  This ring is endowed with the usual polynomial addition operation, but the multiplication, which is associative and distributes over addition, is generally non-commutative, being characterized by the property that $x a = \sigma(a) x$ for every $a \in \Fqm$. For a vector $\vb=(b_0,\ldots, b_{n-1})\in\Fqm^n$, let $\Vand_n(\vb)\in\Fqm^{n\times n}$ be the $\sigma$-Vandermonde matrix as defined in \cite[Def. 2]{liu2023linearized}. 
  The set $\Omega=\{b_0,\ldots, b_{n-1}\}\subseteq \Fqm$ is a \emph{P-independent set} if, and only if, $\rk(\Vand_n(\vb))=n$~\cite[Lem.~12]{lam1988vandermonde}. 
  The following code family, which is based on $\sigma$-Vandermonde matrices and  generalizes Reed--Solomon and Gabidulin codes \cite{reed1960polynomial, gabidulin1985theory}, was introduced in \cite{liu2015construction,martinez-penas2017skew}. %
  
  \begin{definition}[\textbf{Linearized Reed--Solomon (LRS) Codes}] \label{def:LRSC}
  Let~$n=rg\geq k$ and $\Fqm$ be such that $1\leq g\leq q-1$ and $r\leq m$ hold. 
  Choose $\va=(a_1,a_1,\ldots,a_{g})\in \Fqm^g$ such that all elements $a_i$ are from different conjugacy classes of $\Fqm$. 
  Let $\vbeta=\left(\vbeta_1,\vbeta_2,\ldots,\vbeta_{g}\right)\in \Fqm^n$ where the entries in each $\vbeta_i\in \Fqm^{r}$, $i\in [g]$ are $\Fq$-linearly independent and let $\vb\in\Fqm^n$ with $b_\mu=a_{i(\mu)} (\beta^{(j(\mu))}_{i(\mu)})^{q-1}$ for $\mu\in[0,n-1]$.
     An $(n,k)$ linearized Reed--Solomon code over $\Fqm$ on $(\vb,\vbeta)$ with respect to $\sigma$ is given by 
      \begin{equation*}
          \Clrs(\vb,\vbeta):=\left\{ f(\vb)\odot \vbeta \mid f\in \Fqmx, \deg(f)<k \right\}
      \end{equation*}
      with $f(\vb)=(f(b_0),\ldots,f(b_{n-1}))\in \Fqm^n$.
  \end{definition}

  \subsection{Skew Lagrange Polynomials and Lagrange Basis}
  \label{subsec:skew-lagrange}
  Lagrange-type skew polynomials 
  can be constructed by Newton interpolation 
  for skew polynomials 
  \cite[Prop.~2.6]{martinez-penas2022codes}.
  
  \begin{definition}[\textbf{Skew Lagrange Polynomials}]\label{def:skew-lagrange-poly}
      Let $\Omega=\{a_0,a_1,\ldots,a_{k-1}\}\subseteq \Fqm$ be a P-independent set. 
      A skew Lagrange polynomial $\ell_i^{\Omega}\in \Fqmx$ fulfills the constraints $\ell_i^{\Omega}(a_i)=1$ and $\ell_i^{\Omega}(a_j)=0$ for all $i,j\in[0,k-1], j\neq i$. 
  \end{definition}
  
  Given a skew polynomial $f\in\Fqmx$ of degree $k-1$ evaluated on a P-independent set $\Omega$, it can also be written in a form with the Lagrange basis $(\ell_0^{\Omega},\ldots, \ell_{k-1}^{\Omega})\in(\Fqmx)^k$, instead of the form with the monomial basis $(1,x,\dots,x^{k-1})\in(\Fqmx)^k$. More details on the transformation between the monomial and the Lagrange bases are given in Appendix \ref{appendix:monomial-laglange}.

  \subsection{Maximally Recoverable Locally Repairable Codes}
  \label{subsec:MRLRC}
\begin{definition}[\text{\textbf{MR-LRC} \cite{blaum2013partial,gopalan2014explicit}}]
\label{def:MR-LRC}
An LRC $\Cod\subseteq \fiel^n$ with $g$ groups and local distance $\delta_i$ for $i\in[g]$ is said to be \emph{maximally recoverable}, i.e., MR-LRC, if 
after puncturing at most $\delta_i-1$ positions
in each group, the 
punctured code is still MDS.
\end{definition}

  The MR-LRC construction used throughout this work is taken from \cite{martinez-penas2019universal} and defined next for completeness. It uses LRS codes and has been proven to be an MR-LRC in \cite[Th.~2]{martinez-penas2019universal}.
  
  \begin{construction}[\hspace{-1pt}{\cite[Constr.~1]{martinez-penas2019universal}}]\label{con:MR-LRC_linearized}
      Let $g$ be the number of local groups with equal locality $r$ and local distance $\delta$. Choose $\Fqm$ such that $m \geq r$ and $q > \max(g,r+\delta-2)$.
      The construction of the code has two steps:
      \begin{enumerate}
          \item \emph{Outer code}: Choose an $(n,k)$ LRS code  $\Cout \subseteq \Fqm^n$ for $n=rg$.
          \item \emph{Local codes}: Choose any $(r+\delta-1,r)$ MDS code $\Cloci\subseteq \Fq^{r+\delta-1}$ which is linear over the local field $\Fq$ for $i\in[g]$.
      \end{enumerate}
      The global code $\Cglob\subseteq \Fqm^N$ with $N=n+g (\delta-1)=g (r+\delta-1)$ is then defined by 
      \begin{equation*}
          \Cglob=\{\vc\out\cdot \diag\left(\vA_1,\vA_2,\ldots,\vA_{g}\right) \mid \vc\out\in\Cout\}, 
      \end{equation*}
      with $\vA_i\in \Fq^{r\times (r+\delta-1)}$ being the generator matrix of $\Cloci$ for $i\in [g]$. The number of global parities of this construction is $h=n-k=rg-k$.
  \end{construction}

  To be able to securely store data in DSSs encoded by MR-LRCs in the presence of an $(l_1,l_2)$-eavesdropper, we use the following construction, which is similar to the construction from \cite[Th. 33]{rawat2014optimal} based on Gabidulin codes. 
  \begin{construction}
  \label{con:secure_lrc_linearized} 
      Let $\ke$ be the number of independent symbols that an $(l_1,l_2)$-eavesdropper observes in a DSS storing $k$ independent symbols and $\ks=k-\ke$.
      Assume $\ks>0$. The construction has the following steps:
      \begin{enumerate}
          \item Given a file $\vu\hs=(u_0,u_1,\ldots,u_{\ks-1})$ composed of $\ks$ symbols in $\Fqm$,
          generate $\vr=(r_0,r_1,\ldots,r_{\ke-1})$, where the symbols $r_i$ are independent and uniformly distributed over $\Fqm$, for all $i\in[0,\ke-1]$.
          Append $\vr$ to $\vu\hs$ to obtain $\vu=(\vr,\vu\hs)\in\Fqm^{k}$.
          \item Encode 
          $\vu$ by the MR-LRC as in \cref{con:MR-LRC_linearized}.
      \end{enumerate}
  \end{construction}
  As we will show in \cref{subsec:upper_bounds}, with the two novel global repair schemes introduced in \cref{sec:direct-forwarded}, this construction has positive secrecy dimensions.

  \section{Novel Global Repair of MR-LRCs}
  \label{sec:global-rep}
We first present a naive global repair scheme for MR-LRCs and point out that its secrecy dimension is zero. We then introduce a framework which allows for positive secrecy dimension in a DSS encoded by an MR-LRC in the presence of an $(l_1,l_2)$-eavesdropper.
  \subsection{The Naive Global Repair of MR-LRC}
  \label{sec:naive-global-repair}
  Consider a DSS encoded by \cref{con:secure_lrc_linearized}. 
  In a naive global repair, each failed node downloads as many symbols from other nodes as needed for its own recovery.
  Note that a global repair is needed only if more than $\delta-1$ nodes fail in some group. In this case, the nodes in this group need to download $k-\nu$ symbols from other groups to reconstruct their data, where $\nu$ is the number of intact nodes in their group.  
  It follows from the definition of MR-LRCs that after puncturing the $\delta-1$ failed nodes, the code is still an MDS code. Hence, as long as the failed nodes gather any $k$ symbols from the intact nodes, they can reconstruct the whole data stored in the DSS (see also \cite[Sec.III]{martinez-penas2019universal}).
  
  The drawback of such a naive global repair is that if an $(l_1,l_2)$-eavesdropper with $l_2 > 0$ were to observe the nodes that require a global repair, it can also reconstruct the whole DSS after observing the global repair, leading to zero secrecy dimension.

   \subsection{Direct and Forwarded Global Repair}
  \label{sec:direct-forwarded}

  In our proposed global repair schemes, we assume that each group has a central processing unit (CPU) which coordinates the global repair process.
  If a global repair is needed in a group, the CPU of this group sends a request to the other CPUs.
 The CPU of each group, having sufficiently many intact nodes, collects the symbols from its group needed for the global repair and summarizes them into one symbol. 
  This symbol is then sent to the CPU of the group that needs the global repair.
    
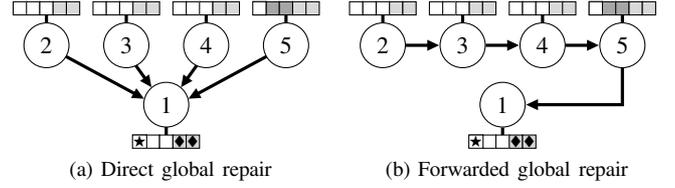
\begin{figure} [t]
      \centering
      \subfloat[Direct global repair\label{fig:direct}]{
          \resizebox{0.47\columnwidth}{!}{\begin{tikzpicture}[
          square/.style = {draw, rectangle, minimum size=\dn, outer sep=0, inner sep=0, font=\small},
        cpu/.style = {circle, draw, minimum size = \m, font=\small, fill=white, outer sep=0},
                                  ]
          \def\m{15pt}
          \def\w{5}
          \def\h{4}
          \def\dn{5pt}
          \def\nodeoff{14pt}
          \def\loc{3}

          \node [cpu]  (g2) at (-3*\m,0) {2};
          \node [cpu]  (g3) at (-1*\m,0) {3};
          \node [cpu]  (g4) at (1*\m,0) {4};
          \node [cpu]  (g) at (3*\m,0) {5};
  
          \node [cpu]  (g1) at (0,-1.5*\m) {1};

          \foreach \x in {1,...,\w}
          \foreach \y in {1,...,\h}
             {  
                \ifnum\x>\loc
                        \node [square, fill=gray!30]  (\x_\y) at (-5*\m+2*\y*\m+\x*\dn-3*\dn,\nodeoff) {};
                 \else
                    \ifnum\y>3
                        \node [square, fill=gray!70]  (\x_\y) at (-5*\m+2*\y*\m+\x*\dn-3*\dn,\nodeoff) {};
                    \else
                        \node [square, fill=white]  (\x_\y) at (-5*\m+2*\y*\m+\x*\dn-3*\dn,\nodeoff) {};
                    \fi
                \fi
             }

          \foreach \x in {1,...,\w}
             {  
                \ifnum\x>\loc
                        \node [square, fill=gray!30]  (\x_5) at (\x*\dn-3*\dn,-1*\nodeoff-1.5*\m) {\tiny $\blacklozenge$};
                 \else
                        \node [square, fill=white]  (\x_5) at (\x*\dn-3*\dn,-1*\nodeoff-1.5*\m) {};
                \fi
             }

             \node [square, fill=white]  (node_5-1) at (-5*\m+8*\m-2*\dn,\nodeoff) {};

            \node [square, fill=white]  (node_1-1) at (-2*\dn,-1*\nodeoff-1.5*\m) {\tiny $\bigstar$};

          \draw[very thick] (g2.north)--(3_1.south);
          \draw[very thick] (g3.north)--(3_2.south);
          \draw[very thick] (g4.north)--(3_3.south);
          \draw[very thick] (g.north)--(3_4.south);
          \draw[very thick] (g1.south)--(3_5.north);

              \draw[-{Triangle[angle=45:5pt]}, very thick] (g2.330) -- (g1.165);
              \draw[-{Triangle[angle=45:5pt]}, very thick] (g3.295) -- (g1.130);
              \draw[-{Triangle[angle=45:5pt]}, very thick] (g4.245) -- (g1.50);
              \draw[-{Triangle[angle=45:5pt]}, very thick] (g.210) -- (g1.15);

      \end{tikzpicture}}
      } \hfil
  \subfloat[Forwarded global repair\label{fig:forwarded}]{
      \resizebox{0.47\columnwidth}{!}{\begin{tikzpicture}[
          square/.style = {draw, rectangle, minimum size=\dn, outer sep=0, inner sep=0, font=\small},
        cpu/.style = {circle, draw, minimum size = \m, font=\small, fill=white},
                                  ]
          \def\m{15pt}
          \def\w{5}
          \def\h{4}
          \def\dn{5pt}
          \def\nodeoff{14pt}
          \def\loc{3}

          \node [cpu]  (g2) at (-3*\m,0) {2};
          \node [cpu]  (g3) at (-1*\m,0) {3};
          \node [cpu]  (g4) at (1*\m,0) {4};
          \node [cpu]  (g) at (3*\m,0) {5};
  
          \node [cpu]  (g1) at (0,-1.5*\m) {1};

          \foreach \x in {1,...,\w}
          \foreach \y in {1,...,\h}
             {  
                \ifnum\x>\loc
                        \node [square, fill=gray!30]  (\x_\y) at (-5*\m+2*\y*\m+\x*\dn-3*\dn,\nodeoff) {};
                 \else
                    \ifnum\y>3
                        \node [square, fill=gray!70]  (\x_\y) at (-5*\m+2*\y*\m+\x*\dn-3*\dn,\nodeoff) {};
                    \else
                        \node [square, fill=white]  (\x_\y) at (-5*\m+2*\y*\m+\x*\dn-3*\dn,\nodeoff) {};
                    \fi
                \fi
             }

        \foreach \x in {1,...,\w}
             {  
                \ifnum\x>\loc
                        \node [square, fill=gray!30]  (\x_5) at (\x*\dn-3*\dn,-1*\nodeoff-1.5*\m) {\tiny $\blacklozenge$};
                 \else
                        \node [square, fill=white]  (\x_5) at (\x*\dn-3*\dn,-1*\nodeoff-1.5*\m) {};
                \fi
             }

             \node [square, fill=white]  (node_5-1) at (-5*\m+8*\m-2*\dn,\nodeoff) {};

            \node [square, fill=white]  (node_1-1) at (-2*\dn,-1*\nodeoff-1.5*\m) {\tiny $\bigstar$};

          \draw[very thick] (g2.north)--(3_1.south);
          \draw[very thick] (g3.north)--(3_2.south);
          \draw[very thick] (g4.north)--(3_3.south);
          \draw[very thick] (g.north)--(3_4.south);
          \draw[very thick] (g1.south)--(3_5.north);
  
          \draw[-{Triangle[angle=45:5pt]}, very thick] (g2.east) -- (g3.west);
          \draw[-{Triangle[angle=45:5pt]}, very thick] (g3.east) -- (g4.west);
          \draw[-{Triangle[angle=45:5pt]}, very thick] (g4.east) -- (g.west);
          \draw[-{Triangle[angle=45:5pt]}, very thick] (g.south) |- (g1.east);
          
  \end{tikzpicture}}
  }
      \caption{Illustration of two different global repair schemes where an erasure (star) in the first group is repaired with global repair. Each circle depicts a CPU of a group that coordinates a repair. The nodes in the groups are depicted as the little squares. The forwarding list for (b) is $\FL=\{2,3,4,5,1\}$ }
      \label{fig:MR-LRC_global_repair_schemes}
  \end{figure}

  We introduce two schemes for the global repair process: direct repair and forwarded repair, which only differ in the manner in which each CPU sends its contribution to the CPU of the group needing a global repair.
  In the direct global repair scheme, the CPUs of the intact groups send their contribution
  directly to the CPU of the group that needs a repair.
  In the  forwarded global repair scheme, 
  each CPU of the intact groups
  forwards its contribution to the next CPU according to a forwarding list $\FL$. 
  At each CPU, the new contribution is the sum of its own contribution
  and, if applicable, the received contribution from the previous CPU in the forwarding list. 
  Therefore, this scheme can be seen as an aggregate-and-forward scheme.
  In the forwarded repair scheme, each group receives at most one symbol. Hence, potentially increasing the secrecy dimension compared to the direct repair. 
  The two schemes are illustrated in \cref{fig:MR-LRC_global_repair_schemes} with five groups and a global repair required in the first group. 

  \subsection{Local Polynomials}
  \label{sec:local-poly}
  We introduce in the following the key tool which enables each CPU to send at most one symbol to the group where the global repair is needed. This tool relies on the use of outer LRS codes in \cref{con:MR-LRC_linearized}.
  
  We define a \emph{minimal global repair set} as a set $\Delglob\subseteq\{(i,j) \mid i\in[g], j\in[0,r+\delta-2])\}$ of intact nodes storing independent symbols and $|\Delglob|=k$. For a fixed group $s\in [g]$, it holds that $|\Delglob\cap \{(s,j)\mid j\in[0,r+\delta-2]\}|\leq r$.  
  In \cref{fig:MR-LRC_example}, we have $\Delglob=\{(1,1),(2,0),(2,3),(2,4),(3,1),(3,2),(3,3)\}$.

  \begin{definition}[\textbf{Local Polynomial}]\label{def:local_polynomial}
      Let $\Cod\glob\subseteq \Fqm^{N}$ and $\Cout\subseteq \Fqm^{n}$  be the global and outer code from \cref{con:MR-LRC_linearized}. 
      Fix a minimal global repair set $\Delglob$ of nodes.
      For a codeword $\vc=(c_1^{(0)},c_1^{(1)},\dots, c_{g}^{(r+\delta-2)})\in\Cglob$,
      the \emph{local polynomial} $L_i\in\Fqm[x;\sigma]$ of the $i$-th group has the following properties:
   \begin{itemize}
      \item $L_i(\tilde{b}_i^{(j)})=c_i^{(j)}/\tilde{\beta}_i^{(j)}$ for all $(i,j)\in \Delglob$,
      \item $L_i(\tilde{b}_s^{(t)})=0$ for all $s\neq i$ and $(s,t)\in \Delglob$,
   \end{itemize}  
   where $\tilde{\beta}_i^{(j)}$  are entries in $\tilde{\vbeta}=\vbeta\diag\left(\vA_1,\vA_2,\ldots,\vA_{g}\right)$ and $\tilde{b}_i^{(j)}=a_{i} (\tilde{\beta}^{(j)}_{i})^{q-1}$.
   
  \end{definition}
Similar to the skew Lagrange polynomials, the local polynomials can be generated by the Newton interpolation~\cite[Prop.~2.6]{martinez-penas2022codes}.

  \begin{theorem}\label{th:local_polynomial_sum}
  We follow the notations in \cref{def:local_polynomial}.
      Let $f$ be an encoding polynomial of the outer code $\Cout=\Clrs(\vb,\vbeta)$. 
      Let $\Delglobone:=\{i\mid (i,j)\in \Delglob\}$.  
      It holds that 
      \begin{equation}\label{eq:sum_local_poly}
          f=\sum_{i\in \Delglobone} L_i.
      \end{equation}
  \end{theorem}
  \begin{IEEEproof}
      See Appendix \ref{appendix:proof-thm-local-poly-sum}.
  \end{IEEEproof}

  \section{Secrecy Dimension of MR-LRCs}
  \label{sec:sec-cap}
  This section investigates the secrecy dimension of a DSS encoded by \cref{con:secure_lrc_linearized} with a direct or forwarded repair scheme.
        The following lemmas are needed to show the secrecy of \cref{con:secure_lrc_linearized}.
    \begin{lemma}[\textbf{Secrecy Lemma}{\cite{shah2011information}}]\label{lem:Secrecy}
    Consider a DSS storing $\vu=(\vu\hs,\vr)$ as in \cref{con:secure_lrc_linearized}.
    Let $\RvU\hs$, $\RvR$ and $\RvE$ be the random variables corresponding to $\vu\hs$, $\vr$ and the symbols observed by an eavesdropper, respectively.
      If $\Ent(\RvE)\leq \Ent(\RvR)$ and $\Ent(\RvR\mid \RvU\hs,\RvE)=0$, then 
      the eavesdropper cannot gain any information about $\vu\hs$, i.e., $\MI(\RvU\hs;\RvE)=0$.
  \end{lemma}

  \begin{lemma}\label{lem:entropy_function_bound}
  Let $\RvK=(\RvK_0,\RvK_1,\ldots,\RvK_{k-1})$ be a vector consisting of $k\in \nat$ random variables.
  Consider 
  a vector $\RvX = (\RvX_0,\RvX_1,\ldots,\RvX_{m-1})$ consisting of $m\in \nat$ random variables such that $\RvX= \vM \RvK\trans$, 
  where $\vM$ is of size ${m\times k}$. 
  It holds that $\Ent(\RvX) \leq k$.
  Furthermore, if the variables in $\RvK$ are i.i.d.,
  then $\Ent(\RvX) =\rk(\vM)$. 
  \end{lemma}   
  \begin{IEEEproof}
      See Appendix \ref{appendix:proof-thm-local-poly-sum}.
  \end{IEEEproof}

  We now show that \cref{con:secure_lrc_linearized} is information theoretically secure. 
   \begin{theorem}
   \label{th:glob_sec_cap}
   Consider a DSS storing $\vu=(\vu\hs,\vr)$ as in \cref{con:secure_lrc_linearized}.
   We have
      \begin{equation*}
          \MI(\RvU\hs;\RvE)=0,
      \end{equation*}
      where $\RvU\hs$ and $\RvE$ are the random variables corresponding to the securely stored symbols and the observed symbols by the eavesdropper, respectively.
  \end{theorem}
  \begin{IEEEproof}
  We prove the statement via \cref{lem:Secrecy}.
  The first step is to show that $\Ent(\RvE)\leq \Ent(\RvR)$ is fulfilled. By \cref{con:secure_lrc_linearized}, we choose $\vr$ consisting of $\ke$ random symbols. Thus, the first condition holds.
  
  The second step is to show that $\Ent(\RvR\mid \RvU\hs, \RvE)=0$. 
  The eavesdropped symbols 
  are $\ve = \vM\vc_{\Delglob},$
      where the matrix $\vM\in\Fqm^{(\ke\times k)}$ represents the $\ke$ independent symbols that an eavesdropper has as constraints on the $k$ encoded symbols as in \cref{con:secure_lrc_linearized}.  
  The basis is the outer codeword at the global repair set without the column multipliers of the LRS code, i.e., $\vc_{\Delglob}:=(\vc\out\odot \vbeta^{-1})|_{\Delglob}$.
      The matrix $\vM$ can be transformed in the domain of the monomial coefficients of the encoding skew polynomial $f$ by \cref{def:monomial-lagrange-basis} yielding
      \begin{equation*}
          \ve \Vand_{k}(\vb)^{-1} = \underbrace{\vM \Vand_{k}(\vb)^{-1}}_{=:\vT\he} \vf,
      \end{equation*}
      where $\vf$ denotes the coefficients of the encoding skew polynomial $f$ and $\vb$ are the code locators of the LRS code.
      The matrix $\vT\he$ gives $\ke$ independent constraints on the polynomial coefficients $\vf$, since $\vM$ has rank $\ke$ and $\Vand_{k}(\vb)^{-1}$ has full rank. 
      Together with the $\ks$ coefficients of the information symbols, with random vector representation $\RvU\hs$, we have $k=\ke+\ks$ constraints on  $k$ coefficients of $f$. 
      It remains to show that the $k$ constraints are independent. 
      The $\ks$ coefficients can be written in a matrix $\vT\hs$ such that $\vu\hs=\vT\hs\vf$. The stacked matrix $\vT$ consisting of $\vT\he$ and $\vT\hs$  can be used to determine $\Ent(\RvU\hs,\RvE)$ by \cref{lem:entropy_function_bound}.
      The matrix $\vT\hs$ has $\ks$ nonzero entries which are on the main diagonal. They contribute $\ks$ to the rank of $\vT$.
      If the Vandermonde matrix $\Vand_{k}(\vb)^{-1}$ is punctured at the corresponding $\ks$ columns, it still has rank $k-\ks$ due to its structure. 
      The overall rank of the punctured matrix $\vT|_{\ks}$ is therefore still $k-\ks=\ke$.
      Thus, the matrix $\vT$ has rank $k$ and all coefficients of $f$, including the random symbols $\vr$, can be determined given $\ve$ and $\vu\hs$, i.e., $\Ent(\RvR\mid \RvU\hs, \RvE)=0$ holds.
  \end{IEEEproof}
  Thus, for $k-\ke>0$, \cref{con:secure_lrc_linearized} can guarantee a positive secrecy dimension against an $(l_1,l_2)$-eavesdropper that observes $\ke$ independent symbols.

  \subsection{Secrecy Dimension for Direct and Forwarded Global Repair}
  \label{subsec:upper_bounds}
    The next step is to quantify the secrecy dimension of DSS using MR-LRCs with a direct or forwarded global repair.
    It can be calculated by 
    \begin{equation*}
        \ks=\Ent(\RvK\mid\RvE)=\Ent(\RvK)-\Ent(\RvE)=k-\ke,
    \end{equation*} 
    which follows from $\Ent(\RvK,\RvE)=\Ent(\RvK)=k$ due to $\RvE=f(\RvK)$.
    Thus, we quantify the secrecy dimension by calculating  $\Ent(\RvE)$.

\subsubsection{Direct Global Repair}
\label{subsubsec:dir_glob}

  Before the secrecy dimension with the direct global repair scheme in the presence of an $(l_1,l_2)$-eavesdropper is calculated, some preliminary considerations should be made.
  First, note that the eavesdropper only gains knowledge  during global repair if the global repair is performed in a group that is observed in an $l_2$-manner. Direct global repair does not reveal any information to the other groups which are only sending information. 
  Second, if $l_2=0$, the secrecy dimension is $\ks=k-l_1$ since the eavesdropper does not make any observations when global repairs are performed. Therefore, we only consider $l_2\geq 1$.
  Third, the number of globally repairable erasures is bounded from above by the number of global parities $h=gr-k$. If more than $h$ failed nodes need to be globally repaired, part of the data cannot be recovered. 

  \begin{theorem}[\textbf{Secrecy Dimension with Direct Global Repair}]\label{th:direct_glob_rate}
  Consider a DSS encoded by \cref{con:secure_lrc_linearized} with locality $r$, $g$ groups and $h\leq r$ global parities in the presence of an $(l_1,l_2)$-eavesdropper with  $l_2\geq 1$ and $l_1+l_2r<k$. 
      The secrecy dimension of the DSS with a direct global repair is
      \begin{equation}
      \label{eq:sec_rate_dir}
          \ks\dir =  k -\underbrace{\left(l_2 r+l_1 -h + \sum_{i=1}^{g} \min(h,r-e_i) \right)}_{\ke\dir},
      \end{equation}
      where $e_i$ denotes the number of independent symbols that the eavesdropper is observing in the $i$-th group from the $l_2r+l_1$ nodes, i.e., in the static case before the global repair process. 
  \end{theorem}
  \begin{IEEEproof}
      
      The proof is given in Appendix \ref{appendix:sec_rate_proofs}.
  \end{IEEEproof}
  \emph{Remark}: Note that for $h\geq r$, the secrecy dimension will be zero since $\ke\dir=k$. This can be verified by assuming that $h=r$, which means $k=n-h=gr-r$.
   Thus, it holds that $\ke\dir=l_2r+l_1-r+\sum_{i=1}^{g} \min(r,r-e_i)$, where we have by definition of $e_i$ that $\ke\dir = (g-1)r$.

\subsubsection{Forwarded Global Repair}
\label{subsubsec:for_glob}
  Before the secrecy dimension of a DSS with forwarded global repair is stated, we make some preliminary considerations.
  First, note that 
  global repairs required by the groups observed in an $l_2$-manner does not add information to the eavesdropper,
  since the eavesdropper has observed the data stored in these groups before the nodes failed.
  Second, 
  if two groups, that are observed in an $l_2$-manner, are next to each other in the forwarding list. 
  The second group will only receive a symbol that is already known by the eavesdropper.
  Third,
  if the eavesdropper is at the beginning of the forwarding list, it does not receive a symbol and can thus not gain any knowledge during global repair.
  These special cases drive us to derive the secrecy dimension of a DSS with the forwarded global repair only for $g\geq 3$.
  For $g\leq 2$, the secrecy dimension is only determined by the static observations, i.e.,
  $\ks=k-(l_2r+l_1)$.

Denote by $\FL$ the forwarding list of the forwarded scheme and by $\FL\upi$ the forwarding list containing the groups that are upstream with respect to the $i$-th group. Let $\Gtwo$ be the set of groups that are observed in an $l_2$-manner.
If a group in $\FL\upi$ is observed in an $l_2$-manner, let %
$$\FLbar\upi:=\FL\upi \setminus \left\{\bigcup_{\nu\in \FL\upi\cap \Gtwo}\{ j\in \FL\upn{\nu}\}\cup \{\nu\}\right\}.$$
In words, 
given a group that is observed in an $l_2$-manner, 
$\FLbar\upi$ is the set of groups that are upstream in the forwarding list $\FL\upi$ between the $l_2$-observed group with index $i$ and the next $l_2$-observed group (excluded) or until the end of the list (included). %
For example, in \cref{fig:forwarded}, assume the group 3 and 5 were observed in an $l_2$-manner.
Then the two lists are $\FL\upn{5}=\{2,3,4\}$ and $\FLbar\upn{5} =\{4\}$.
  \begin{theorem}[\textbf{Secrecy Dimension with Forwarded Global Repair}]
  \label{th:for_glob_rate} 
  Consider a DSS encoded by \cref{con:secure_lrc_linearized} with locality $r$, $g\geq 3$ groups and $h\leq r$ global parities in the presence of an $(l_1,l_2)$-eavesdropper with $l_2\geq 1$ and $l_1+l_2r<k$.
  
    The  secrecy dimension of forwarded global repair is
      \begin{equation}
      \label{eq:sec_rate_for}
          \ks\for =   k -\underbrace{\left((l_2 r+l_1) + \sum_{i\in\Gtwo}\min(h,\sum_{j\in\FLbar\upi} (r-e_j))\right)}_{\ke\for},
      \end{equation}
      where $e_i$ denotes the number of independent symbols that the eavesdropper is observing in the $i$-th group from the $l_2r+l_1$ nodes, i.e., in the static case before the global repair process.
  \end{theorem}
  \begin{IEEEproof}
      The proof is given in Appendix \ref{appendix:sec_rate_proofs}.
  \end{IEEEproof}

  \subsection{Comparison of Direct and Forwarded Global Repair}
  \label{subsec:comparison}

We give two examples in which we compare the secrecy dimension of the introduced schemes.

\begin{figure} [t]
      \centering
      \subfloat[Direct global repair\label{fig:with-erasure-symbols-direct-repair}]{
          \resizebox{0.47\columnwidth}{!}{\begin{tikzpicture}[
      square/.style = {draw, rectangle, minimum size=\m, outer sep=0, inner sep=0, font=\small},
      cpu/.style = {circle, draw, minimum size = \m, font=\small, fill=white},
                              ]
      \def\m{20pt}
      \def\off{25pt}
      \def\circoff{6pt}
      \def\trioff{7pt}
      \def\tri{6pt}
      \def\circsize{2pt}
      \def\w{5}
      \def\h{3}
      \def\loc{3}

        \foreach \x in {1,...,\w}
          \foreach \y in {1,...,\h}
             {  
             \pgfmathtruncatemacro{\xin}{\x-1}
                \ifnum\x>\loc
                    \ifnum\y<\h
                        \node [square, fill=gray!30]  (\x_\y) at (\x*\m,-\y*\off) {$ \underline{x}^{(\xin)}_{\y} $};
                    \else
                        \node [square, fill=gray!30]  (\x_\y) at (\x*\m,-\y*\off) {$ \underline{x}^{(\xin)}_{\y} $};
                    \fi
                 \else
                    \ifnum\y>2
                        \node [square, fill=gray!70]  (\x_\y) at (\x*\m,-\y*\off) {$ \underline{x}^{(\xin)}_{\y} $};
                    \else
                        \node [square, fill=white]  (\x_\y) at (\x*\m,-\y*\off) {$ \underline{x}^{(\xin)}_{\y} $};
                    \fi
                \fi
             }

      \node [square, fill=gray!30]  (b) at (4*\m,-1*\off) {$\blacklozenge$};
      \node [square, fill=gray!30]  (c) at (5*\m,-1*\off) {$\blacklozenge$};
  
      \node [square, fill=white]  (d) at (2*\m,-2*\off) {$\blacklozenge$};
      \node [square, fill=white]  (e) at (3*\m,-2*\off) {$\blacklozenge$};
  
      \node [square, fill=white]  (f) at (1*\m,-3*\off) {$\blacklozenge$};
      \node [square, fill=gray!30]  (g) at (5*\m,-3*\off) {$\blacklozenge$};
  
      \node [square, fill=white]  (l) at (3*\m,-1*\off) {$\bigstar$};
      \node [square, fill=white]  (l) at (1*\m,-1*\off) {$\bigstar$};
  
       \node[cpu] (cpu1) at (-0.25*\m, -1*\off) {1};

      \node[cpu] (cpu2) at (-0.25*\m, -2*\off) {2};

      \node[cpu] (cpu3) at (-0.25*\m, -3*\off) {3};
  
      \filldraw[violettblau] (1*\m-\circoff,-2*\off+\circoff) circle (\circsize);
      \filldraw[rot] (1*\m-\trioff-\tri,-1*\off+\trioff) -- (1*\m-\trioff,-1*\off+\trioff) -- (1*\m-\trioff-\tri/2,-1*\off+\trioff+\tri) --(1*\m-\trioff-\tri,-1*\off+\trioff);

      \draw (cpu3.180) edge[-{Triangle[angle=45:5pt]}, very thick, in=200, out = 160] (cpu1.160);
      \draw (cpu2.160) edge[-{Triangle[angle=45:5pt]}, very thick, in=220, out = 160] (cpu1.200);

      \draw[very thick] (cpu1.east) -- (1_1.west);
      \draw[very thick] (cpu2.east) -- (1_2.west);
      \draw[very thick] (cpu3.east) -- (1_3.west);
              
\end{tikzpicture}}
      } \hfil
  \subfloat[Forwarded global repair\label{fig:with-erasure-symbols-forwarded-repair}]{
  \resizebox{0.47\columnwidth}{!}{\begin{tikzpicture}[
      square/.style = {draw, rectangle, minimum size=\m, outer sep=0, inner sep=0, font=\small},
                              ]
      \def\m{20pt}
      \def\off{25pt}
      \def\circoff{6pt}
      \def\trioff{7pt}
      \def\tri{6pt}
      \def\circsize{2pt}
      \def\w{5}
      \def\h{3}
      \def\loc{3}

        \foreach \x in {1,...,\w}
          \foreach \y in {1,...,\h}
             {  
             \pgfmathtruncatemacro{\xin}{\x-1}
                \ifnum\x>\loc
                    \ifnum\y<\h
                        \node [square, fill=gray!30]  (\x_\y) at (\x*\m,-\y*\off) {$ \underline{x}^{(\xin)}_{\y} $};
                    \else
                        \node [square, fill=gray!30]  (\x_\y) at (\x*\m,-\y*\off) {$ \underline{x}^{(\xin)}_{\y} $};
                    \fi
                 \else
                    \ifnum\y>2
                        \node [square, fill=gray!70]  (\x_\y) at (\x*\m,-\y*\off) {$ \underline{x}^{(\xin)}_{\y} $};
                    \else
                        \node [square, fill=white]  (\x_\y) at (\x*\m,-\y*\off) {$ \underline{x}^{(\xin)}_{\y} $};
                    \fi
                \fi
             }

      \node [square, fill=gray!30]  (b) at (4*\m,-1*\off) {$\blacklozenge$};
      \node [square, fill=gray!30]  (c) at (5*\m,-1*\off) {$\blacklozenge$};
  
      \node [square, fill=white]  (d) at (2*\m,-2*\off) {$\blacklozenge$};
      \node [square, fill=white]  (e) at (3*\m,-2*\off) {$\blacklozenge$};
  
      \node [square, fill=white]  (f) at (1*\m,-3*\off) {$\blacklozenge$};
      \node [square, fill=gray!30]  (g) at (5*\m,-3*\off) {$\blacklozenge$};

      \node [square, fill=white]  (l) at (3*\m,-1*\off) {$\bigstar$};
      \node [square, fill=white]  (l) at (1*\m,-1*\off) {$\bigstar$};

      \node[circle, draw, minimum size = \m, font=\small, fill=white] (cpu1) at (-0.25*\m, -1*\off) {1};

      \node[circle, draw, minimum size = \m, font=\small, fill=white] (cpu2) at (-0.25*\m, -2*\off) {2};

      \node[circle, draw, minimum size = \m, font=\small, fill=white] (cpu3) at (-0.25*\m, -3*\off) {3};

      \filldraw[violettblau] (1*\m-\circoff,-2*\off+\circoff) circle (\circsize);
      \filldraw[rot] (1*\m-\trioff-\tri,-3*\off+\trioff) -- (1*\m-\trioff,-3*\off+\trioff) -- (1*\m-\trioff-\tri/2,-3*\off+\trioff+\tri) --(1*\m-\trioff-\tri,-3*\off+\trioff);

      \draw (cpu3.200) edge[-{Triangle[angle=45:5pt]}, very thick, in=210, out = 160] (cpu1.200);
      \draw (cpu2.200) edge[-{Triangle[angle=45:5pt]}, very thick, in=140, out = 200] (cpu3.160);

      \draw[very thick] (cpu1.east) -- (1_1.west);
      \draw[very thick] (cpu2.east) -- (1_2.west);
      \draw[very thick] (cpu3.east) -- (1_3.west);
              
      \end{tikzpicture}}
  }
      \caption{Illustration of the global repair schemes for the DSS from \cref{fig:MR-LRC_example}.
      The failed nodes marked with stars need to be repaired globally.
      In (a) they are repaired by direct global repair and in (b) forwarded global repair is used with the forwarding list $\FL=\{2,3,1\}$. Both repairs are coordinated by the CPUs of the groups, depicted by the circles on the left. 
      The secrecy rates of the DSS with respective repair schemes are given in \cref{ex:1}.} 
      \label{fig:MR-LRC_example_erasure_repair}
  \end{figure}
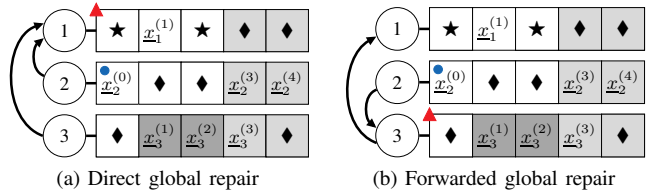

  \begin{example}
  \label{ex:1}
      Consider the DSS depicted in \cref{fig:MR-LRC_example} again. 
      It has $g=3$ groups, locality $r=3$, local distance $\delta=3$ and $h=2$ global parities. The eavesdropper is a $(1,1)$-eavesdropper. 
      The eavesdropped nodes and the failed nodes are depicted in \cref{fig:MR-LRC_example_erasure_repair}.
      With the directed global repair illustrated in \cref{fig:with-erasure-symbols-direct-repair}, 
      we can compute the secrecy dimension from \eqref{eq:sec_rate_dir} in \cref{th:direct_glob_rate}:
      $\ks\dir=k-\ke\dir=7-6=1$. Namely, one information symbol can be stored securely on the considered system with the direct scheme.
        With the forwarded global repair illustrated in \cref{fig:with-erasure-symbols-forwarded-repair}, from \eqref{eq:sec_rate_for} in \cref{th:for_glob_rate},
        we get that $\ks\for=k-\ke\for=7-6=1$ information symbol can be stored securely on the considered system with the forwarded scheme.
         \end{example}

\begin{figure}[t]
	\centering
	\resizebox{0.9\columnwidth}{!}{\begin{tikzpicture}
\begin{axis}[
width = 8cm,
height = 5cm,
	fill=white,
	grid style={dotted,gray},
	xmajorgrids,
	xminorgrids=true,
	ymajorgrids,
	yminorticks=true,
 legend pos=north west,
	legend style={fill,fill opacity=0.75, text opacity=1, draw=none},
    tick align=outside,
    tick pos=left,
    xlabel={\scriptsize number of groups $g$},
    xtick={1,3,...,15},
    xmin=1,
    xmax=15,
    xtick style={color=black},
    ylabel={\scriptsize secrecy dimension $\ks$},
    ymajorgrids,
    ymin=0, 
    ymax=100, %
    ytick style={color=black},
    ticklabel style = {font=\scriptsize}
]

\addplot [rot, dotted, mark = +,  mark options={solid}]
table[col sep=comma]{
1, 0
2, 4
3, 8
4, 12
5, 16
6, 20
7, 24
8, 28
9, 32
10,36
11,40
12, 44
13, 48
14, 52
15, 56
};
\addlegendentry{\scriptsize direct global repair};
\label{plot:direct}

\addplot [violettblau, dotted, mark=x,  mark options={solid}] 
table[col sep=comma]{
1, 0
2, 4
3, 8
4, 15
5, 22
6, 29
7, 36
8, 43
9, 50
10,57
11, 64
12, 71
13, 78
14, 85
15, 92
};
\label{plot:forwarded}
\addlegendentry{\scriptsize  forwarded global repair};

\addplot [black, dotted, mark=star,  mark options={solid}] 
table[col sep=comma]{
1, 0
2, 7
3, 14
4, 21
5, 28
6, 35
7, 42
8, 49
9, 56
10, 63
11, 70
12, 77
13, 84
14, 91
15, 98
};
\label{plot:upper_bound}
\addlegendentry{\scriptsize LRC without global repair};

\end{axis}

\end{tikzpicture}}
	\caption{Plot of the secrecy dimension of a DSS that uses an MR-LRC with forwarded global repair (blue) and  direct global repair (red) for fixed parameters $l_2=1$, $l_1=0$ $r=7$, $h=3$.  The secrecy dimensions are the same for  $g\leq 3$. For $g>3$ forwarded global repair has a higher secrecy dimension. 
    In addition, the secrecy dimension of an LRC-coded DSS without global repair, i.e., $h=0$ is plotted. 
 }
	\label{fig:sec_rate_comp}
\end{figure}
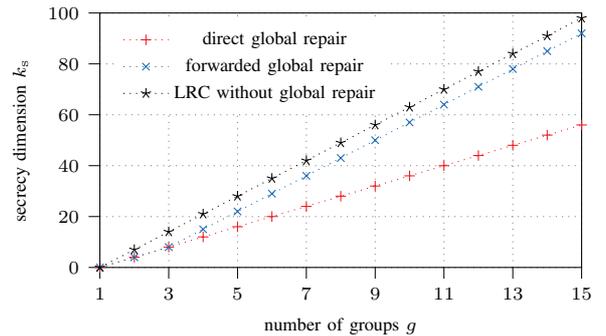

\begin{example}
\label{ex:2}
We now compare the secrecy dimension of the two global repair schemes for different number of groups $g$.
Consider a DSS with locality $r=7$ and $h=3$ global parities in the presence of an $(0,1)$-eavesdropper.
   The DSS stores $k=r\cdot(g-1)+(r-h)=7\cdot(g-1)+4$ symbols. 
   The comparison is presented in \cref{fig:sec_rate_comp}, where the secrecy dimension of a DSS encoded by a conventional LRC with $h=0$ global parity is also plotted.
   We can see that forwarded global repair has a higher secrecy dimension than direct global repair for $g>3$. 
   The secrecy dimension of the forwarding global repair scheme is only slightly below the secrecy dimension of a conventional LRC.
\end{example}

  However, forwarded global repair has the drawback of an increasing latency in $g$, since each group, except the first group in the forwarding list, is waiting for the upstream contribution before sending its contribution.

\section{Conclusion and Outlook}
\label{sec:sum}
We have introduced a new repair framework for MR-LRCs with LRS codes. In the framework, we associate a central processing unit to each local group that uses local polynomials to summarize the global repair contribution from the local group.
Two different global repair schemes are proposed and their secrecy dimensions in the presence of a passive eavesdropper are determined.
For future research, it would be interesting to investigate the secrecy dimension of MR-LRCs for arbitrary repair graphs, i.e., global repair schemes that have an arbitrary global repair topology consisting of forwarding (line) and collecting (tree) structures \cite{patra2022node}.
Moreover, the general secrecy capacity of DSS that use MR-LRCs could be investigated.

\section*{Acknowledgements}
The authors want to thank Prof. Antonia Wachter-Zeh for initializing this collaboration and Prof. Stephan ten Brink for supporting this work. 

\IEEEtriggeratref{16}

\bibliographystyle{IEEEtran}
\bibliography{bibliofile}

\begin{thebibliography}{10}
\providecommand{\url}[1]{#1}
\csname url@samestyle\endcsname
\providecommand{\newblock}{\relax}
\providecommand{\bibinfo}[2]{#2}
\providecommand{\BIBentrySTDinterwordspacing}{\spaceskip=0pt\relax}
\providecommand{\BIBentryALTinterwordstretchfactor}{4}
\providecommand{\BIBentryALTinterwordspacing}{\spaceskip=\fontdimen2\font plus
\BIBentryALTinterwordstretchfactor\fontdimen3\font minus
  \fontdimen4\font\relax}
\providecommand{\BIBforeignlanguage}[2]{{%
\expandafter\ifx\csname l@#1\endcsname\relax
\typeout{** WARNING: IEEEtran.bst: No hyphenation pattern has been}%
\typeout{** loaded for the language `#1'. Using the pattern for}%
\typeout{** the default language instead.}%
\else
\language=\csname l@#1\endcsname
\fi
#2}}
\providecommand{\BIBdecl}{\relax}
\BIBdecl

\bibitem{gopalan2012on}
P.~Gopalan, C.~Huang, H.~Simitci, and S.~Yekhanin, ``On the locality of
  codeword symbols,'' \emph{IEEE Transactions on Information Theory}, vol.~58,
  no.~11, pp. 6925--6934, 2012.

\bibitem{huang2007pyramid}
C.~Huang, M.~Chen, and J.~Li, ``Pyramid {{Codes}}: {{Flexible Schemes}} to
  {{Trade Space}} for {{Access Efficiency}} in {{Reliable Data Storage
  Systems}},'' in \emph{Sixth {{IEEE International Symposium}} on {{Network
  Computing}} and {{Applications}} ({{NCA}} 2007)}, Jul. 2007, pp. 79--86.

\bibitem{papailiopoulos2014locally}
D.~S. Papailiopoulos and A.~G. Dimakis, ``Locally {{Repairable Codes}},''
  \emph{IEEE Transactions on Information Theory}, vol.~60, no.~10, pp.
  5843--5855, Oct. 2014.

\bibitem{kamath2014codes}
G.~M. Kamath, N.~Prakash, V.~Lalitha, and P.~V. Kumar, ``Codes {{With Local
  Regeneration}} and {{Erasure Correction}},'' \emph{IEEE Transactions on
  Information Theory}, vol.~60, no.~8, pp. 4637--4660, Aug. 2014.

\bibitem{gopalan2014explicit}
P.~Gopalan, C.~Huang, B.~Jenkins, and S.~Yekhanin, ``Explicit maximally
  recoverable codes with locality,'' \emph{IEEE Transactions on Information
  Theory}, vol.~60, no.~9, pp. 5245--5256, 2014.

\bibitem{chen2007maximally}
M.~Chen, C.~Huang, and J.~Li, ``On the maximally recoverable property for
  multi-protection group codes,'' in \emph{2007 IEEE International Symposium on
  Information Theory}.\hskip 1em plus 0.5em minus 0.4em\relax IEEE, 2007, pp.
  486--490.

\bibitem{martinez-penas2019universal}
U.~Martínez-Peñas and F.~Kschischang, ``Universal and dynamic locally
  repairable codes with maximal recoverability via sum-rank codes,'' \emph{IEEE
  Transactions on Information Theory}, vol.~PP, pp. 1--1, 06 2019.

\bibitem{gopi2020maximally}
S.~Gopi, V.~Guruswami, and S.~Yekhanin, ``Maximally {{Recoverable LRCs}}: {{A
  Field Size Lower Bound}} and {{Constructions}} for {{Few Heavy Parities}},''
  \emph{IEEE Transactions on Information Theory}, vol.~66, no.~10, pp.
  6066--6083, Oct. 2020.

\bibitem{guruswami2020constructions}
V.~Guruswami, L.~Jin, and C.~Xing, ``Constructions of {{Maximally Recoverable
  Local Reconstruction Codes}} via {{Function Fields}},'' \emph{IEEE
  Transactions on Information Theory}, vol.~66, no.~10, pp. 6133--6143, Oct.
  2020.

\bibitem{cai2022construction}
H.~Cai, Y.~Miao, M.~Schwartz, and X.~Tang, ``A {{Construction}} of {{Maximally
  Recoverable Codes With Order-Optimal Field Size}},'' \emph{IEEE Transactions
  on Information Theory}, vol.~68, no.~1, pp. 204--212, Jan. 2022.

\bibitem{gopi2022improved}
S.~Gopi and V.~Guruswami, ``Improved {{Maximally Recoverable LRCs Using Skew
  Polynomials}},'' \emph{IEEE Transactions on Information Theory}, vol.~68,
  no.~11, pp. 7198--7214, Nov. 2022.

\bibitem{blaum2013partial}
M.~Blaum, J.~L. Hafner, and S.~Hetzler, ``Partial-{MDS} codes and their
  application to raid type of architectures,'' \emph{IEEE Transactions on
  Information Theory}, vol.~59, no.~7, pp. 4510--4519, 2013.

\bibitem{calis2016general}
G.~Calis and O.~O. Koyluoglu, ``A general construction for {PMDS} codes,''
  \emph{IEEE Communications Letters}, vol.~21, no.~3, pp. 452--455, 2016.

\bibitem{gabrys2018constructions}
R.~Gabrys, E.~Yaakobi, M.~Blaum, and P.~H. Siegel, ``Constructions of partial
  {MDS} codes over small fields,'' \emph{IEEE Transactions on Information
  Theory}, vol.~65, no.~6, pp. 3692--3701, 2018.

\bibitem{neri2020random}
A.~Neri and A.-L. {Horlemann-Trautmann}, ``Random construction of partial
  {{MDS}} codes,'' \emph{Designs, Codes and Cryptography}, vol.~88, no.~4, pp.
  711--725, Apr. 2020.

\bibitem{bogart2021constructing}
T.~Bogart, A.-L. {Horlemann-Trautmann}, D.~Karpuk, A.~Neri, and M.~Velasco,
  ``Constructing {{Partial MDS Codes}} from {{Reducible Algebraic Curves}},''
  \emph{SIAM Journal on Discrete Mathematics}, vol.~35, no.~4, pp. 2946--2970,
  Jan. 2021.

\bibitem{rawat2014optimal}
A.~S. Rawat, O.~O. Koyluoglu, N.~Silberstein, and S.~Vishwanath, ``Optimal
  locally repairable and secure codes for distributed storage systems,''
  \emph{IEEE Transactions on Information Theory}, vol.~60, no.~1, pp. 212--236,
  2014.

\bibitem{agarwal2016security}
A.~Agarwal and A.~Mazumdar, ``Security in locally repairable storage,''
  \emph{IEEE Transactions on Information Theory}, vol.~62, no.~11, pp.
  6204--6217, 2016.

\bibitem{liu2023linearized}
H.~Liu, H.~Wei, A.~Wachter-Zeh, and M.~Schwartz, ``Linearized reed-solomon
  codes with support-constrained generator matrix,'' in \emph{2023 IEEE
  Information Theory Workshop (ITW)}.\hskip 1em plus 0.5em minus 0.4em\relax
  IEEE, 2023, pp. 7--12.

\bibitem{lam1988vandermonde}
\BIBentryALTinterwordspacing
T.~Lam and A.~Leroy, ``Vandermonde and wronskian matrices over division
  rings,'' \emph{Journal of Algebra}, vol. 119, no.~2, pp. 308--336, 1988.
  [Online]. Available:
  \url{https://www.sciencedirect.com/science/article/pii/0021869388900634}
\BIBentrySTDinterwordspacing

\bibitem{reed1960polynomial}
\BIBentryALTinterwordspacing
I.~S. Reed and G.~Solomon, ``Polynomial codes over certain finite fields,''
  \emph{Journal of the Society for Industrial and Applied Mathematics}, vol.~8,
  no.~2, pp. 300--304, 1960. [Online]. Available:
  \url{https://doi.org/10.1137/0108018}
\BIBentrySTDinterwordspacing

\bibitem{gabidulin1985theory}
E.~Gabidulin, ``Theory of codes with maximum rank distance (translation),''
  \emph{Problems of Information Transmission}, vol.~21, pp. 1--12, 01 1985.

\bibitem{liu2015construction}
S.~Liu, F.~Manganiello, and F.~R. Kschischang, ``Construction and decoding of
  generalized skew-evaluation codes,'' in \emph{2015 IEEE 14th Canadian
  Workshop on Information Theory (CWIT)}.\hskip 1em plus 0.5em minus
  0.4em\relax IEEE, 2015, pp. 9--13.

\bibitem{martinez-penas2017skew}
U.~Martínez-Peñas, ``Skew and linearized {Reed-Solomon} codes and maximum sum
  rank distance codes over any division ring,'' \emph{Journal of Algebra}, vol.
  504, 10 2017.

\bibitem{martinez-penas2022codes}
\BIBentryALTinterwordspacing
U.~Martínez-Peñas, M.~Shehadeh, and F.~R. Kschischang, ``Codes in the
  sum-rank metric: Fundamentals and applications,'' \emph{Foundations and
  Trends® in Communications and Information Theory}, vol.~19, no.~5, pp.
  814--1031, 2022. [Online]. Available:
  \url{http://dx.doi.org/10.1561/0100000120}
\BIBentrySTDinterwordspacing

\bibitem{shah2011information}
N.~B. Shah, K.~V. Rashmi, and P.~V. Kumar, ``Information-theoretically secure
  regenerating codes for distributed storage,'' in \emph{2011 IEEE Global
  Telecommunications Conference - GLOBECOM 2011}, 2011, pp. 1--5.

\bibitem{patra2022node}
A.~Patra and A.~Barg, ``Node repair on connected graphs,'' \emph{IEEE
  Transactions on Information Theory}, vol.~68, no.~5, pp. 3081--3095, 2022.

\bibitem{gander2005change}
\BIBentryALTinterwordspacing
W.~Gander, ``Change of basis in polynomial interpolation,'' \emph{Numerical
  Linear Algebra with Applications}, vol.~12, no.~8, pp. 769--778, 2005.
  [Online]. Available:
  \url{https://onlinelibrary.wiley.com/doi/abs/10.1002/nla.450}
\BIBentrySTDinterwordspacing

\end{thebibliography}

\clearpage
\appendices
\crefname{appsec}{Appendix}{Appendices}
\section{Monomial and Lagrange Basis of Skew Polynomials}
\label{appendix:monomial-laglange}

  \begin{definition}[\textbf{Monomial and Lagrange Basis}]\label{def:monomial-lagrange-basis}\\
      Let $f=f_0+f_1 x+\ldots+f_{k-1} x^{k-1}\in \Fqmx$ be a skew polynomial 
      in monomial basis with coefficient vector $\vf=(f_0,f_1,\ldots,f_{k-1})\in \Fqm^{k}$. 
      Let $\Omega=\{a_0,a_1,\ldots,a_{k-1}\}\subseteq \Fqm$ be a P-independent set 
      and $\Phi=\{p_0,p_1,\ldots,p_{k-1}\}\subseteq \Fqm$ the set of evaluations of $f$ such that $f(a_i)=p_i$ for all $i\in[0,k-1]$.
      Denote $\vp=(p_0,p_1,\ldots,p_{k-1})\in \Fqm^{k}$. 
      Let $\mathcal{L}=\{\ell_{0},\ell_{1},\ldots,\ell_{k-1}\}$ be a Lagrange basis on $\Omega$ as defined in \cref{def:skew-lagrange-poly}. 
      A skew polynomial $f$ can then be written as 
      $f=p_0\ell_0+p_1\ell_1+\ldots+p_{k-1}\ell_{k-1}$, i.e., it has two representations 
         $f=\vf \cdot\vm(x)=\vp\cdot\vell(x),$ 
      where $\vm(x)=(1,x,\ldots,x^{k-1})\trans$ and $\vell(x)=(\ell_0,\ell_1,\ldots,\ell_{k-1})\trans$.
    \end{definition}
    \begin{lemma}
    \label{lem:lagrange-monomial}
    The two representations from \cref{def:monomial-lagrange-basis} of a skew polynomial $f$  have the $k\times k$ skew Vandermonde matrix $\Vand_k(\va)\in\Fqm^{k\times k}$ with $\va=(a_0,a_1,\ldots,a_{k-1})\in \Fqm^k$ as the transformation matrix. It holds that $\vf \Vand_k(\va)=\vp$,
      and $\vm(x) = \Vand_k(\va) \vell(x)$.
    \end{lemma}
    The proof is by definition of the two basis and the skew Vandermonde matrix and can be found in \cite{gander2005change} for conventional polynomials.

\section{Proof of Theorem \ref{th:local_polynomial_sum}}
\label{appendix:proof-thm-local-poly-sum}

\begin{IEEEproof}
From \cite[Prop.~2.9]{martinez-penas2022codes}, we know that multiplying the outer code $\Cout$ from the right with a block diagonal matrix $\vA\in\Fq^{n\times t}$ can also be realized by adjusting the elements of $\vbeta\in\Fqm^n$ such that $\tilde{\vbeta}=\vbeta\cdot\vA \in \Fqm^t$. We take at most $r$ global codeword symbols $c_i^{(j)}$ from each group. Therefore, the corresponding $\tilde{\beta}_i^{(j)}$ are also $\Fq$-linearly independent since the generator matrices $\vA\in\Fq^{r\times(r+\delta-1)}$ are MDS.
    The sum of local polynomials is equivalent to the encoding polynomial if their evaluations at $k$ points are the same. We can consider
    \begin{equation*}
        f(\tilde{b}_i^{(j)})\tilde{\beta}_i^{(j)}=c_i^{(j)} \text{ for all } (i,j)\in \Delglob
    \end{equation*}
    at $k$ positions.
    For a fixed $(i,j)\in\Delglob$, we have
    \begin{equation*}
    \begin{aligned}
                 f(\tilde{b}_i^{(j)})\tilde{\beta}_i^{(j)} &= \sum_{m\in\Delglobone} L_m(\tilde{b}_i^{(j)})\tilde{\beta}_i^{(j)} \\
                 &= L_i(\tilde{b}_i^{(j)})\tilde{\beta}_i^{(j)} + \sum_{\substack{m\in \Delglobone\\ m\neq i}} L_m(\tilde{b}_i^{(j)})\tilde{\beta}_i^{(j)}\\
                \overset{(a)}&{=} c_i^{(j)} + 0 = c_i^{(j)},
    \end{aligned}
    \end{equation*}
    where $(a)$ holds by \cref{def:local_polynomial}.
   Since $\Delglob$ has cardinality $k$, the sum of local polynomials is equal to $f$ at $k$ P-independent points and thus \eqref{eq:sum_local_poly} holds.
\end{IEEEproof}

\section*{Proof of \cref{lem:entropy_function_bound}}
    \begin{IEEEproof}
        We know that 
        \begin{equation*}
            \Ent(\RvK,\RvX) = \Ent(\RvX) +\Ent(\RvK\mid\RvX)
        \end{equation*}
        by the chain rule of entropy and thus it holds that 
        \begin{equation*}
            \Ent(\RvX) \leq \Ent(\RvK,\RvX)
        \end{equation*}
        with equality if, and only if, $\RvX$ essentially determines $\RvK$, i.e.,~$\Ent(\RvK\mid\RvX)=0$.\\
        Furthermore, it holds that $\RvX=f(\RvK)=\vM\RvK\trans$ which yields
        \begin{equation*}
            \Ent(\RvK,\RvX) = \Ent(\RvK,\vM\RvK\trans)=\Ent(\RvK)\leq k.
        \end{equation*}
        As a result,
        \begin{equation*}
           \Ent(\RvX)=\Ent(\vM\RvK\trans) \leq \Ent(\RvK) \leq k
        \end{equation*}
        holds.
        For the second part, we know that $\Ent(\RvK) = k$ since the random vector $\RvK$ consists of uniformly and independent distributed random variables. The entropy of $\RvX$ is
        \begin{equation*}
            \Ent(\RvX)= \Ent(\vM\RvK\trans)\leq k.
        \end{equation*}
        The rank of the matrix $\vM$ determines how many symbols of the random vector  $\RvX$ are independent and this can be expressed by $\Ent(\RvX) =\rk(\vM)$.    
    \end{IEEEproof}

  \section{Calculation of Secrecy Dimensions}
\label{appendix:sec_rate_proofs}
The following lemma is used to determine the secrecy dimensions with a direct and forwarded global repair.
  \begin{lemma}\label{lem:matrix_structure}
       Let $f\in \Fqmx$ be a skew polynomial of degree $k-1$ in monomial basis with coefficient vector $\vf=(f_0,f_1,\ldots,f_{k-1})\in \Fqm^{k}$. Let $\Omega=\{a_0,a_1,\ldots,a_{n-1}\}\subseteq \Fqm$ be a P-independent set.
      Split the sets $\Omega$ into two subsets $\Omega_k=\{a_i\mid i\in [0,k-1]\}$ and $\Omega_{n-k}=\{a_i\mid i\in [k,n-1]\}$.
      Let $\mathcal{L}=\{\ell_{0},\ell_{1},\ldots,\ell_{k-1}\}$ be a Lagrange basis on $\Omega_k$.
      The matrix 
      \begin{equation*}
          \vM=\begin{pmatrix}
              \ell_{0}^{\Omega_k}(a_{k})&\ell_{1}^{\Omega_k}(a_{k})&\cdots&\ell_{k-1}^{\Omega_k}(a_{k})\\
              \ell_{0}^{\Omega_k}(a_{k+1})&\ell_{1}^{\Omega_k}(a_{k+1})&\cdots&\ell_{k-1}^{\Omega_k}(a_{k+1})\\
              \vdots&\vdots&\ddots&\vdots\\
              \ell_{0}^{\Omega_k}(a_{n-1})&\ell_{1}^{\Omega_k}(a_{n-1})&\cdots&\ell_{k-1}^{\Omega_k}(a_{n-1})
          \end{pmatrix}
      \end{equation*}
      can be written as $\vM = \left(\Vand_k(\va_k)^{-1}\Vand_k(\va_d) \right)\trans$ with $\va_k=(a_0,\dots,a_{k-1})$ and $\va_d=(a_k,\dots,a_{n-1})$. Moreover, $\vM$ has full rank, i.e., $\rk(\vM)=\min(k,d)$.
  \end{lemma}
  \begin{IEEEproof}
      It can be shown that the matrix $\vM$ has full rank by decomposing $\vM\trans$, which has the same rank as $\vM$, into several matrices that are proven to have full rank. It holds that
      \begin{equation*}
          \vM\trans=\vL\Vand_k(\va_d)
      \end{equation*}
      with entries in $\vL$ being the coefficient vectors of the Lagrange skew polynomials $\ell_{i}^{\Omega_k},i\in[0,k-1]$, i.e., $L_{i,j}=\ell_{i,j}^{\Omega_k}$,
      where $\ell_{i,j}^{\Omega_k}$ is the $j$-th coefficient of the $i$-th polynomial $\ell_i^{\Omega_k}$.
      By \cref{lem:lagrange-monomial}, we know that $\vL=\Vand_k(\va_k)^{-1}$ 
      since it holds that $\vell(x)=\vm(x)\vL$.
      Overall, we have $\vM\trans=\Vand_k(\va_k)^{-1}\Vand_k(\va_d) $.
      Both matrices have full rank and it holds that $\rk(\vM)=\rk(\vM\trans)=\min(k,d)$.    
  \end{IEEEproof}
  \emph{Remark}: The above lemma also implies that submatrices of $\vM$ have full rank since they are also a product of two Vandermonde matrices.

\section*{Proof of \cref{th:direct_glob_rate}}

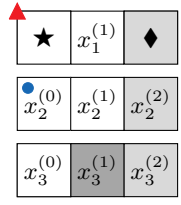
\begin{figure} [t]
      \centering   
      \begin{tikzpicture}[
      square/.style = {draw, rectangle, minimum size=\m, outer sep=0, inner sep=0, font=\small},
                              ]
      \def\m{20pt}
      \def\off{25pt}
      \def\circoff{6pt}
      \def\trioff{7pt}
      \def\tri{6pt}
      \def\circsize{2pt}
      \def\w{3}
      \def\h{3}
      \def\loc{2}

        \foreach \x in {1,...,\w}
          \foreach \y in {1,...,\h}
             {  
             \pgfmathtruncatemacro{\xin}{\x-1}
                \ifnum\x>\loc
                    \ifnum\y<\h
                        \node [square, fill=gray!30]  (\x_\y) at (\x*\m,-\y*\off) {$x^{(\xin)}_{\y} $};
                    \else
                        \node [square, fill=gray!30]  (\x_\y) at (\x*\m,-\y*\off) {$ x^{(\xin)}_{\y} $};
                    \fi
                 \else
                    \ifnum\y>2
                        \node [square, fill=gray!70]  (\x_\y) at (\x*\m,-\y*\off) {$ x^{(\xin)}_{\y} $};
                    \else
                        \node [square, fill=white]  (\x_\y) at (\x*\m,-\y*\off) {$ x^{(\xin)}_{\y} $};
                    \fi
                \fi
             }

      \node [square, fill=gray!30]  (a) at (3*\m,-1*\off) {$\blacklozenge$};

      \node [square, fill=white]  (d) at (1*\m,-1*\off) {$\bigstar$};

        \node [square, fill=white]  (e) at (1*\m,-3*\off) {$x^{(0)}_{3} $};
  
      \filldraw[violettblau] (1*\m-\circoff,-2*\off+\circoff) circle (\circsize);
      \filldraw[rot] (1*\m-\trioff-\tri,-1*\off+\trioff) -- (1*\m-\trioff,-1*\off+\trioff) -- (1*\m-\trioff-\tri/2,-1*\off+\trioff+\tri) --(1*\m-\trioff-\tri,-1*\off+\trioff);
        
\end{tikzpicture}
      \caption{Illustration of a DSS with $N = 9$ nodes and storing $k=5$ independent symbols. 
      The DSS is encoded by an MR-LRC with $g=3$ groups, locality $r=2$, local distance $\delta=2$ (parities in light gray) and $h=1$ global parities (in dark gray). 
      The DSS is observed by a  $(1,1)$-eavesdropper who can read the downloaded and stored data of any node in the top group (marked by a red triangle) and the data stored on one node (marked by a blue circle). }
     
      \label{fig:MR-LRC_proof}
  \end{figure}

Before we give a general proof, the proof idea is illustrated with an example.
Consider the DSS as depicted in \cref{fig:MR-LRC_proof}. 
The global repair set for $x_1^{(0)}$ is $\Delglob=\{(1,1),(2,0),(2,1),(3,0),(3,1)\}$.
The static observations of the eavesdropper $\ve_{\st}$ before the global repair can be summarized in a matrix $\vM_\st$, where the basis is the outer codeword at the global repair set without the column multipliers of the LRS code, i.e., $\vc_{\Delglob}:=(\vc\out\odot \vbeta^{-1})|_{\Delglob}= (c_{\mathrm{out},i}^{(j)}(\beta_i^{(j)})^{-1})_{(i,j)\in\Delglob}$. For the DSS in \cref{fig:MR-LRC_proof}, we have
\begin{equation*}
       \vM_{\st} = \begin{pmatrix}
          \ell_{1,1}^{\Delglob}(b^\prime) & \ell_{2,0}^{\Delglob}(b^\prime) &\ell_{2,1}^{\Delglob}(b^\prime) &\ell_{3,0}^{\Delglob}(b^\prime) & \ell_{3,1}^{\Delglob}(b^\prime)\\
          1 & 0 & 0 & 0 & 0 \\
          0 & 1 & 0 & 0 & 0 
      \end{pmatrix}
\end{equation*}
with $b^\prime = b_1^{(0)}$ and $\ve_{\st}= \vM_{\st}\vc_{\Delglob}$.
The eavesdropper can also observe the downloaded symbols to recover $x_1^{(0)}$. They can be summarized by $\ve_{\dl}=\vM_{\dl}\vc_{\Delglob}$, where $\vM_{\dl}$ is 
\begin{equation*}
       \vM_{\dl} = \begin{pmatrix}
          \ell_{1,1}^{\Delglob}(b^\prime) &  0 & 0 & 0 & 0\\
          0& \ell_{2,0}^{\Delglob}(b^\prime) &\ell_{2,1}^{\Delglob}(b^\prime) & 0 & 0\\
          0 & 0 & 0 &\ell_{3,0}^{\Delglob}(b^\prime) & \ell_{3,1}^{\Delglob}(b^\prime)
      \end{pmatrix}.  
\end{equation*}
By \cref{lem:entropy_function_bound}, the knowledge of the eavesdropper can be determined by calculating the rank of the stacked matrix $\vM$, which consists of $\vM_{\st}$ and $\vM_{\dl}$, i.e., 
\begin{equation*}
       \vM = \begin{pmatrix}
            \ell_{1,1}^{\Delglob}(b^\prime) & \ell_{2,0}^{\Delglob}(b^\prime) &\ell_{2,1}^{\Delglob}(b^\prime) &\ell_{3,0}^{\Delglob}(b^\prime) & \ell_{3,1}^{\Delglob}(b^\prime)\\
          1 & 0 & 0 & 0 & 0 \\
          0 & 1 & 0 & 0 & 0 \\
          \ell_{1,1}^{\Delglob}(b^\prime) &  0 & 0 & 0 & 0\\
          0& \ell_{2,0}^{\Delglob}(b^\prime) &\ell_{2,1}^{\Delglob}(b^\prime) & 0 & 0\\
          0 & 0 & 0 &\ell_{3,0}^{\Delglob}(b^\prime) & \ell_{3,1}^{\Delglob}(b^\prime)
      \end{pmatrix}.  
\end{equation*}
The first row of $\vM$ is linearly dependent on the last three rows. Thus, $\rk(\vM)=\rk(\vM^\prime)$ with 
\begin{equation*}
       \vM^\prime = \begin{pmatrix}
          1 & 0 & 0 & 0 & 0 \\
          0 & 1 & 0 & 0 & 0 \\
          \ell_{1,1}^{\Delglob}(b^\prime) &  0 & 0 & 0 & 0\\
          0& \ell_{2,0}^{\Delglob}(b^\prime) &\ell_{2,1}^{\Delglob}(b^\prime) & 0 & 0\\
          0 & 0 & 0 &\ell_{3,0}^{\Delglob}(b^\prime) & \ell_{3,1}^{\Delglob}(b^\prime)
      \end{pmatrix}.  
\end{equation*}
With proper row operations on $\vM^\prime$, we have $\rk(\vM^\prime)=\rk(\vM^{\prime\prime})$, where
\begin{equation*}
       \vM^{\prime\prime} = \begin{pmatrix}
          1 & 0 & 0&0 & 0 & 0 \\
          0 & 1 & 0 &0& 0 & 0 \\
          0 &  0 & 0&0 & 0 & 0\\
          0& 0 &0&\ell_{2,1}^{\Delglob}(b^\prime) & 0 & 0\\
          0 & 0 &0& 0 &\ell_{3,0}^{\Delglob}(b^\prime) & \ell_{3,1}^{\Delglob}(b^\prime)
      \end{pmatrix}.  
\end{equation*}
It can be readily seen that $\rk(\vM^{\prime\prime})=2+\rk(\vM^{\prime\prime\prime})$, where
\begin{equation*}
       \vM^{\prime\prime\prime}
       = \begin{pmatrix}
           \ell_{2,1}^{\Delglob}(b^\prime) & 0 & 0\\
           0 &\ell_{3,0}^{\Delglob}(b^\prime) & \ell_{3,1}^{\Delglob}(b^\prime)
      \end{pmatrix}.  
\end{equation*}
Overall, we have that $\ke=\Ent(\RvE)=\rk(\vM)=2+\rk(\vM^{\prime\prime\prime}
)=2+2=4$. Thus, the secrecy dimension of the DSS shown in \cref{fig:MR-LRC_proof} is $\ks=k-\ke=5-4=1$.

We now turn to the proof of \cref{th:direct_glob_rate}.

\begin{IEEEproof}  
Denote by $\Ero$ (w.r.t.~$\Ert^{\st}$)
the set of independent nodes that are observed by an eavesdropper in an $l_1$-(w.r.t.~$l_2$-) manner
in the static case without global repair, respectively.
Without loss of generality, assume that the first $l_2$ groups are observed by the eavesdropper in the $l_2$-manner.
Consider the worst case, there are $h$ failed nodes that need to be globally repaired and they are, without loss of generality, in the first $l_2$ group at the first $h$ positions with $j\in[0,h-1]$.
  We determine the entropy of the eavesdropped symbols $\Ent(\RvE)$ with \cref{lem:entropy_function_bound}. 
  Assume that all possible local repairs are performed such that only erasures in the first group are left.
  By the definition of MR-LRCs (or PMDS codes), whenever a global repair is required, $|\Delglob|=k$. In the following, we denote $\Delglob=\{(i_1,j_1),\cdots,(i_k,j_k)\}$.
  Let the global repair set $\Delglob$ 
      overlap as much as possible with the static  eavesdropper observations $\Ero$ and $\Ert^{\st}$.
  The eavesdropped symbols $\ve$ can be represented by 
  $\ve = \vM \vc_{\Delglob}$, where
  $\vc_{\Delglob}:=(\vc\out\odot \vbeta^{-1})|_{\Delglob}$ is the outer codeword at the global repair set without the column multipliers of the LRS code, and $\vM=\begin{pmatrix}
      \vM_{\st}\\\vM_{\dl} \end{pmatrix}$. 
      The symbols in $\ve_{\st}=\vM_{\st}\vc_{\Delglob}$ are
  the observed stored symbols of which $h$ are being repaired using the global repair set $\Delglob$, and the matrix $\vM_{\st}\in \Fqm^{(l_2 r + l_1)\times k}$ can be expressed as
  \begin{equation}
  \label{eq:M-st-direct}
      \vM_{\st} = \begin{pmatrix}
          \ell_{i_1,j_1}^{\Delglob}(b_1^{(0)}) &\cdots & \ell_{i_k,j_k}^{\Delglob}(b_1^{(0)})\\
          \vdots&\ddots&\vdots\\
          \ell_{i_1,j_1}^{\Delglob}(b_1^{(h-1)}) &\cdots & \ell_{i_k,j_k}^{\Delglob}(b_1^{(h-1)})\\
          &&\\
          & \widetilde{\vI}_{\st}&  \\
          &&
      \end{pmatrix},  
  \end{equation}
  where the columns indexed by $(\Ero\cup\Ert^{\st})\cap\Delglob$ of $\widetilde{\vI}_{\st}$ form an identity matrix. 
  The symbols observed by the eavesdropper during the global repair process are $\ve_{\dl}=\vM_{\dl}\vc_{\Delglob}$, where

  \begin{equation*}
         \vM_{\dl} =    \begin{pNiceMatrix}
      \ell^{\Delglob}_{i_1,j_1}(b_1^{(0)})&\dots&0\\
                  &\vdots&\\
                  0&\dots&\ell^{\Delglob}_{i_k,j_k}(b_1^{(0)})\\
                  &\vdots&\\
                  \ell^{\Delglob}_{i_1,j_1}(b_1^{(h-1)})&\dots&0\\
                  &\vdots&\\
                  0&\dots&\ell^{\Delglob}_{i_k,j_k}(b_1^{(h-1)})
                  \CodeAfter
                  \SubMatrix{.}{1-1}{3-3}{\}}[xshift=2mm, code = {\begin{tikzpicture}\node[align=center] (c) at (1,1.2) {Repair for \\ first symbol };
                \end{tikzpicture}}]
                \SubMatrix{.}{5-1}{7-3}{\}}[xshift=2mm, code = {\begin{tikzpicture}\node[align=center] (c) at (1,-1.1) {Repair for \\ last symbol };
                \end{tikzpicture}}]
    \end{pNiceMatrix}\quad\quad\quad\hphantom{asdf}
  \end{equation*}
  
  It can be seen that the first $h$ rows of $\vM_{\st}$ are linearly dependent on $\vM_{\dl}$. Hence, $\rk(\vM)=\rk(\vM^\prime)$, where $\vM^\prime=\begin{pmatrix}
      \widetilde{\vI}_{\st}\\\vM_{\dl}
  \end{pmatrix}$ and $\rk(\vM^\prime)=\rk(\widetilde{\vI}_{\st})+\rk(\vM_{\dl}|_{\Delglob\setminus(\Ero\cup\Ert^{\st})})$.

  Consider the matrix $\vM_{\dl}|_{\Delglob\setminus(\Ero\cup\Ert^{\st})})$ groupwise for the $i$-th group. 
  If the $i$-th group is fully punctured by the entries of $\widetilde{\vI}_{\st}$, we have $e_i=r$ and there is no nonzero row in $\vM_{\dl}|_{\Delglob\setminus(\Ero\cup\Ert^{\st})}$ corresponding to the $i$-th group. 
  Otherwise, there are still $r-e_i$ columns corresponding to the $i$-th group. The rows corresponding to the $i$-th group are of the structure investigated in \cref{lem:matrix_structure}. 
  They can be represented by the product of two Vandermonde matrices since they correspond to evaluations of the same polynomial
  at P-independent points. 
  Therefore, the contribution of the $i$-th group is $\min(h,r-e_i)$. 
  It holds that 
      \begin{equation*}
      \rk(\vM_{\dl}|_{\Delglob\setminus(\Ero\cup\Ert^{\st})})=\sum_{i=1}^{g} \min(h,r-e_i)
      \end{equation*}
       and we have 
      \begin{equation*}
          \rk(\vM)=l_2r+l_1 -h+ \sum_{i=1}^{g} \min(h,r-e_i),
      \end{equation*}
      which gives us $\Ent(\RvE)$ by \cref{lem:entropy_function_bound}.
  \end{IEEEproof}

  \section*{Proof of \cref{th:for_glob_rate}}

  \begin{IEEEproof}
   The proof is done in a similar manner as for \cref{th:direct_glob_rate}.
   We follow the notations for $\Ero$, $\Ert^{\st}$ and $\Delglob$ as in the proof of \cref{th:direct_glob_rate}.
   Let the $l_1$-observations of the eavesdropper be distributed in such a way that $|\Ero\cap \Delglob|=l_1$.
   Without loss of generality, assume that the global erasures occur in the first group.
   Consider the worst case that the first group is not observed by the eavesdropper in the $l_2$-manner. 
The eavesdropped symbols can be represented by $\ve = \vM \vc_{\Delglob}$ where $\vc_{\Delglob}:=(\vc\out\odot \vbeta^{-1})|_{\Delglob}$ and 
$\vM=\begin{pmatrix}
    \vM_{\st}\\\vM_{\dl}
\end{pmatrix}\in \Fqm^{(l_2r+l_1+l_2 h)\times k}$.
  The matrix $\vM_{\st}\in \Fqm^{(l_2 r + l_1)\times k}$ represents the eavesdropper's observation in the static case, i.e., before the global repair, and
  $\vM_{\st}=\widetilde{\vI}_{\st}$, where 
  the columns indexed by 
  $(\Ero\cup\Ert^{\st})\cap\Delglob$ of $\widetilde{\vI}_{\st}$ form an identity matrix.
  Hence, it contributes $l_2 r + l_1$ to the rank of $\vM$.
      The other part of $\vM$, namely $\vM_{\dl}\in \Fqm^{l_2  h\times k}$ summarizes the global repair symbols that are observed by the eavesdropper,
        \begin{equation*}
      \vM_{\dl} = \begin{pmatrix}
                  \sum_{\nu\in \FL\upn{\Gtwo(1)}} \bell^{\Delglob}_\nu(b_1^{(0)})\\
                  \vdots\\
                                    \sum_{\nu\in \FL\upn{\Gtwo(l_2)}} \bell^{\Delglob}_\nu(b_1^{(h-1)})\\
                  \end{pmatrix}\ ,
  \end{equation*}
  where $\bell^{\Delglob}_\nu(b)=(\ell^{\Delglob}_{\nu;i_1,j_1}(b),\dots,\ell^{\Delglob}_{\nu;i_k,j_k}(b))$  
  with $\ell^{\Delglob}_{\nu;i,j}(b)=\ell^{\Delglob}_{i,j}(b)$ for $\nu=i$ and $\ell^{\Delglob}_{\nu;i,j}(b)=0$ otherwise, for all $(i,j)\in\Delglob$.
  For each repair and each group observed in an $l_2$-manner, one symbol, as the sum of all upstream symbols, is observed. 
  Therefore, there are $l_2h$ rows in $\vM_{\dl}$.
 The matrix $\vM_{\dl}$ can be reduced by Gaussian elimination to a matrix $\vM_{\dl}^{\prime}$ with  $k-(l_2 r + l_1)$ nonzero columns by subtracting the rows of $\vM_{\st}$.
 Moreover, the matrix $\vM_{\dl}^{\prime}$ can be further reduced to matrix $\vM_{\dl}^{\prime\prime}$ such that only the columns corresponding to groups in $\FLbar\upi$ are nonzero. 
      The submatrices of $\vM^{\prime\prime}$ for each $l_2$-manner observed group (only consider the nonzero columns)  %
      have a structure as described in \cref{lem:matrix_structure} and they are of size $h\times (\sum_{j\in\FLbar\upi} (r-e_j))$ with full rank. 
      Thus, the rank of matrix $\vM$ is 
      \begin{align*}
          \rk(\vM)&=\rk\begin{pmatrix}
          \widetilde{\vI}_{\st}\\\vM_{\dl}^{\prime\prime}
      \end{pmatrix}\\
      &=\left((l_2 r+l_1) + \sum_{i\in\Gtwo}\min(h,\sum_{j\in\FLbar\upi} (r-e_j)) \right)\ ,
      \end{align*}
      which gives us $\Ent(\RvE)$ by \cref{lem:entropy_function_bound}.

  \end{IEEEproof}

\end{document}